\documentclass[runningheads]{llncs}
\usepackage{graphicx}
\usepackage{booktabs}
\usepackage[table]{xcolor}
\begin{document}
%

\title{Social media self-branding and success:\\[0.5ex]
\large{Quantitative evidence from a model competition}}
\titlerunning{Can social media activity drive success?}
%
\author{Fabian Braesemann, Fabian Stephany\\[1ex]
December 15, 2019\\[1ex]
Contact: fabian.braesemann@gmx.de}
\authorrunning{Braesemann, Stephany}
%
\institute{}
\maketitle              
\begin{abstract}
Thanks to the availability of large online data sets, it has become possible to quantify success in different fields of human endeavour. The study presented here contributes to this literature in evaluating the effect of social media activity, as a means of 'self-branding', to increase the chances of models being elected for the \textit{Playboy} Magazine's \textit{Playmate of the Year} award. We hypothesise that candidates who actively manage their Instagram accounts can increase their likelihood to win the award: they use social media to gain more followers, who then might vote for them in the award polls. The findings indicate that social media activity actually has predictive capacity to estimate the outcome of the award. We find evidence that candidates who manage their social media accounts more actively than other candidates have a higher probability to become Playmate of the Year. The findings underline the benefits of social media self-branding as a driver of popularity and success.

\end{abstract}
\section{Introduction}
Big online data is being used to quantify success in different fields of human endeavour  \cite{fraiberger_quantifying_2018,sinatra_quantifying_2016,yin_quantifying_2019}. This fast growing strand of research aims to identify and to quantify concepts related to success, such as performance \cite{pappalardo_quantifying_2018} or luck \cite{janosov_success_2020}. Contributing to this emerging \textit{science of success} literature, we investigate whether active social media management by individuals, known as 'self-branding', is associated with success. To investigate this question, we focus on the case of a model competition that might be facilitated by the influence of popularity gained through social media self-branding: the \textit{Playboy} magazine's \textit{Playmate of the Year} award\footnote{This study examines the \textit{Playmate of the Year} competition as it allows to investigate the effects of social media self-branding within the emerging field of \textit{science of success}. We abstain from making any normative contribution to the ongoing relevant and critical discussion \cite{pitzulo_bachelors_2011} on the content or history of the \textit{Playboy} magazine or the competition itself.}. 

We collect measures of social media activity and popularity from the candidates' Instagram profiles who were participating in the Playmate of the Year polls in the German and American editions of the magazine from 2015 to 2019. Using this data set, we perform a logistic regression to identify variables that are associated with higher odds of winning the award. We find that those candidates with more actively managed social media profiles indeed have a higher liklihood of winning. The out-of-sample prediction performance of the statistical model is assessed using cross-validation. The model shows high predictive performance: in three out of eight historical competitions, the model correctly predicts the winner, in three other cases, the eventual winner was predicted second. Overall, the model performs much better than the benchmark of random drawing. We apply the statistical model to predict the outcome of the German 2020 Playmate of the Year award.\footnote{There is no Playmate of the Year award in the United States in 2020: \scriptsize{\url{https://pagesix.com/2020/03/10/playboy-to-end-playmate-of-the-year-in-favor-of-inclusivity/}}.} 
In summary, the study provides a case study to assess the potential value of using social media self-branding in driving popularity and success in the structured context of a model competition.
\vspace{-2ex}
\section{Related Work}
\vspace{-1ex}
Thanks to the availability of large online data sets, it has become possible to quantify success in many fields. For example, Fraiberger et al. (2018) model the career paths of artists based on their initial success in exhibiting their work in established galleries \cite{fraiberger_quantifying_2018}, Sinatra et al. (2016) quantify the evolution of individual scientific impact \cite{sinatra_quantifying_2016}, while Yin et al. (2019) provide a predictive model to understand the time-sensitive patterns that discriminate failure and success in repeated trials \cite{yin_quantifying_2019}.

Success determinants have also been investigated in areas of popular culture, such as sports. For example, Maimone and Yasseri (2019) predict the outcomes of matches in major European football leagues \cite{maimone_football_2019}, and Pappalado and Cintia (2018) investigate the relation between performance and success in football \cite{pappalardo_quantifying_2018}. Other areas of popular culture, that have been investigated from a \textit{science of success} perspective are movie box office success \cite{mestyan_early_2013,bhattacharjee_identifying_2017,ruus_predicting_2019}, music \cite{ginsburgh_expert_2003}, creative careers \cite{janosov_success_2020}, and show business \cite{williams_quantifying_2019}.

The Playboy magazine itself has been investigated as a phenomenon of popular culture. Coulter (2014)  investigates the influence of the magazine on redefining male consumer culture \cite{coulter_selling_2014}. Fraterrigo (2009) explores how the magazine constructed ideals of masculinity \cite{fraterrigo_playboy_2009}; similar studies have been conducted by Osgerby (2001) and Pitzulo (2011) \cite{osgerby_playboys_2001,pitzulo_bachelors_2011}.
These studies investigate the Playboy as an icon of popular culture and discuss it from a sociological and anthropological point of view.

To our knowledge, model contests or similar competitions that are facilitated by popularity have, so far, not been investigated from a quantitative perspective. 

\vspace{-2ex}
\section{Hypotheses}
\vspace{-1ex}
We consider the case of the Playboy magazine's Playmate of the Year award. The nature of the competition allows for a quantification of social media self-branding, online popularity, and success.

First, the number of potential candidates is limited to the twelve \textit{Playmates of the Month} in a given year who have been previously nominated by the magazine's editorial team in each country edition of the magazine. Secondly, the time-frame and organisation of the challenge are consistent across contests in different years and some national editions of the magazine. Thus, candidates know, when being nominated as Playmate of the Month that they will take part in the Playmate of the Year contest. The winners of the award might be able to become successful models or celebrities. Thus, the award provides incentives for the candidates to win. The candidates face a clear outcome variable and it can be assumed that they will aim to increase their chances of winning in gaining a fan base of followers who might vote for them. 

Thirdly, due to the long time frame of the contest (the contest is usually awarded in the spring of the following year) the candidates are likely to make use of social media channels as a means of 'self-branding' \cite{khamis_self_2017,marwick_tweet_2011,senft_microcelebrity_2013} in order to gain a fan base. The candidates might therefore choose to actively contribute to their channels on the photo-sharing social network site Instagram (a popular platform among internet celebrities \cite{djafarova_exploring_2017,marwick_instafame_2015}) in order to build online popularity and to advertise themselves prior to the election.


We hypothesise that the active management of the social media channel of candidates running for the \textit{Playmate of the Year} competition provides them with a larger follower base on their social media profile. The popularity gained by such activity, then, positively affects the likelihood to be elected:
\vspace{-1ex}
\begin{quote}
$H1$\textit{: Online popularity, captured by measures from the candidates social media profile, is positively associated with the likelihood to win the 'Playmate of the Year' award.}
\end{quote}

\vspace{-3ex}
\section{Data \& Methods}
\vspace{-2ex}

To test the research hypothesis, we collect data published by the candidates on their Instagram profiles. As a first step, we identified all Playmates of the Month in the 2015 to 2019 editions of the Playboy Magazine in the United States and Germany. Then, we searched for the Instagram channels of these Playmates. While not all Playmates of the Month maintain an Instagram channel, the majority does so, providing first evidence for the importance of the platform for self-branding of Playboy models. Having identified the Instagram channel, we apply web scraping to get the list of all picture URLs uploaded by the candidates. Then, we collect the date and user name of all commentators of the pictures.\footnote{Neither do we collect the images themselves, nor any content-specific information of the commentators. The user names are solely collected to count the number of unique followers and are not displayed anywhere in the study. Furthermore, we want to emphasise that we do not publish the list of Instagram user names of the candidates; the results presented here exclusively mention the candidates artist's name as published in the Playboy Magazine. No Instagram data is published.}
Based on the data set, we extract a number of features that are then used in the statistical model: (1) an indicator variable measuring if a candidate has won the Playmate of the Year Award, (2) an indicator variable measuring if a candidate is from the country of the Playboy edition, in which the candidate is being nominated as Playmate of the Month, (3) an indicator variable measuring if a candidate has been Playmate of the Month in an earlier edition of the magazine, (4) the number of unique users commenting photos in the time between appearance in the magazine as Playmate of the Month and the election to Playmate of the Year (logarithmic scale), (5) the number of pictures uploaded in the time between appearance in the magazine as Playmate of the Month and the election to Playmate of the Year (logarithmic scale), and (6) the number of comments made by the candidate on her own pictures in the time between appearance in the magazine as Playmate of the Month and the election to Playmate of the Year (logarithmic scale).

Using this parsimonious set of features, we model the odds to become \textit{Playmate of the Year} using a simple logistic regression. In summary, we obtained social media data from 76 candidates who became Playmate of the Month in the American or German edition of the magazine from 2015 to 2018, on which the model is trained. We then extrapolate on the twelve candidates who compete for the 2020 award in the German \emph{Playboy} magazine.

\vspace{-2ex}
 
\section{Results}

\vspace{-2ex}

\begin{figure}[h!]
\centering
\includegraphics[width = 0.8\linewidth]{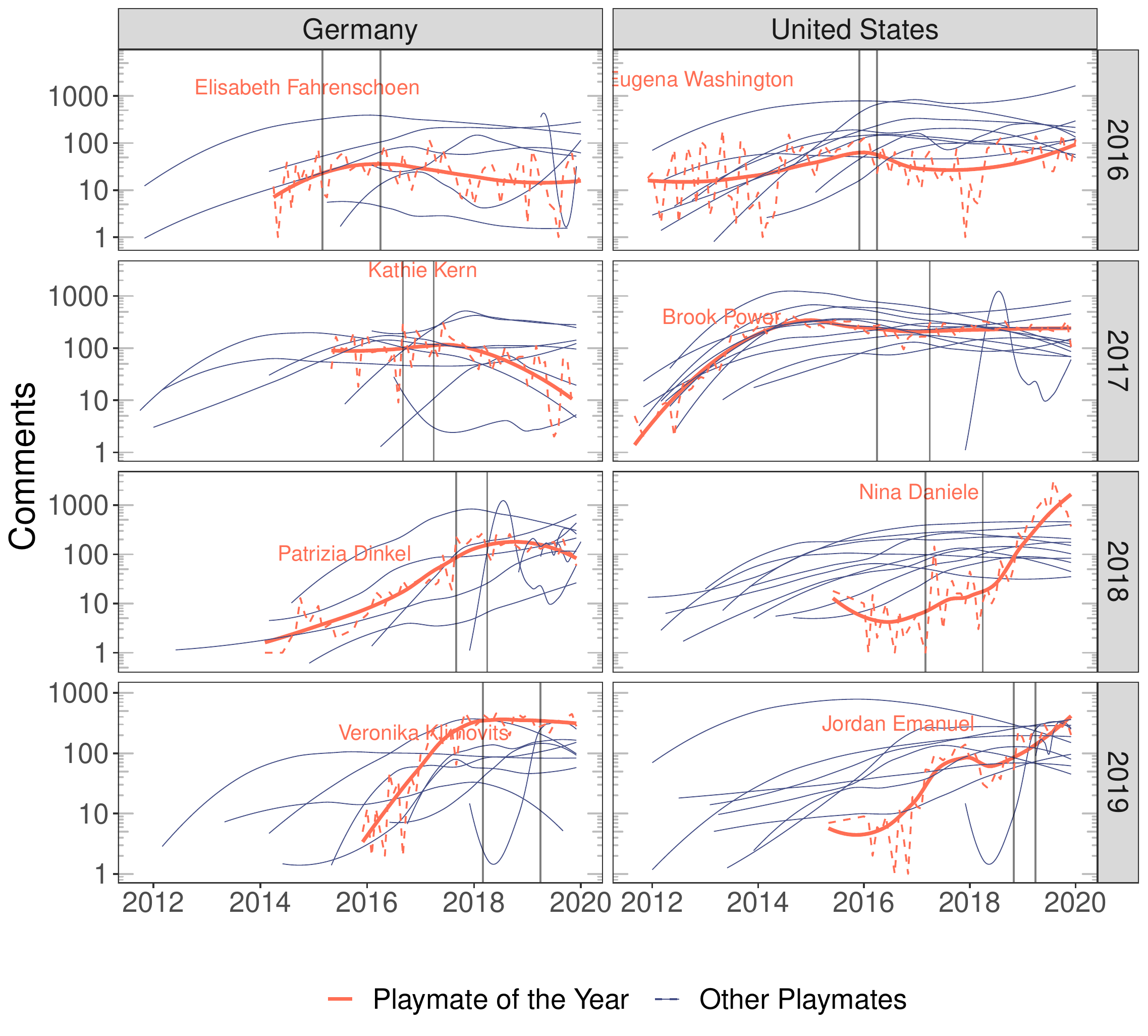}
\caption{\sf{Time-sensitive patterns of Instagram popularity of the \textit{Playmates of the Month} in different competitions.}}
\label{fig:1}
\end{figure}

Figure\,\ref{fig:1} provides descriptive support for the first research hypothesis. It shows the time-sensitive patterns of comments to the Instagram accounts of the candidates in the different contests. The successful candidates are highlighted in red. The successful Playmates were able to attract a relatively large follower base already before their appearance in the magazine (first vertical line). 

\begin{figure}[h!]
\centering
\includegraphics[width = 0.75\linewidth]{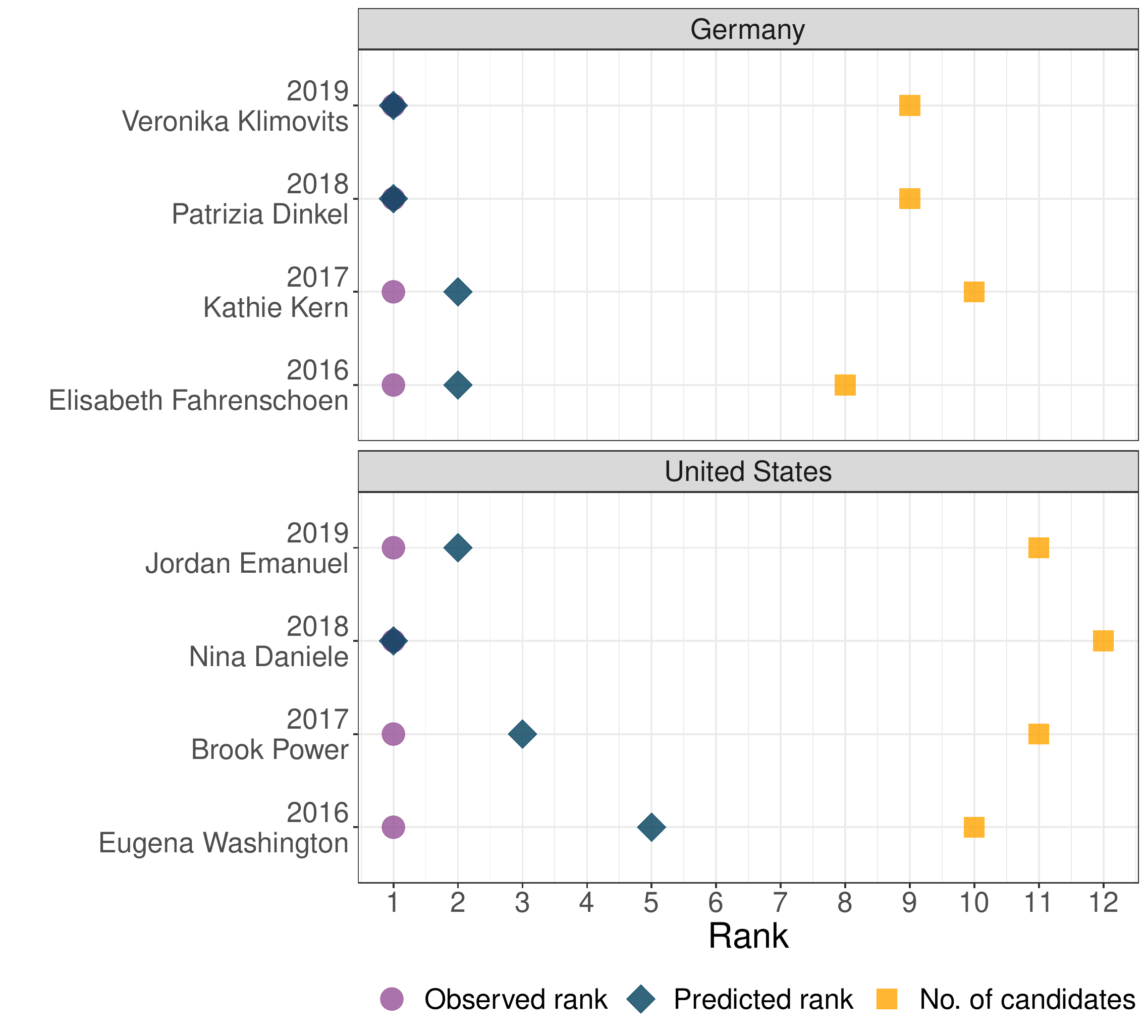}
\caption{\sf{Cross-validated out-of-sample rank prediction of the Playmate of the Year in the US and German editions of the magazine.}}
\label{fig:2}
\end{figure}

Turning to the second research hypothesis, Fig.\,\ref{fig:2} presents the cross-validated out-of-sample rank prediction of the successful candidates. Our final sample is limited to eight competitions, four in the United States and four in Germany, with a maximum of twelve possible candidates. We iteratively train our model with the data from seven competitions while predicting the outcome of the remaining contest. In seven out of eight cases, the model ranks the eventual successful candidate among the top-three (in six cases among the top-two, and correctly predicts three competition outcomes).

These findings clearly show that the model beats the baseline distribution of a simple random draw.\footnote{If the Playmate of the Year was a random pick from a set of twelve competing candidates, one would expect a more uniform distribution of predicted ranks with a winning probability of 8.3\%.} 
We conclude that the social media data has, indeed, predictive power to forecast the outcome of the contest.

Turning to extrapolation, we provide predictions on the outcome of the \textit{2020 Playmate of the Year} contest (Table\,\ref{tab:1}). The highest chances of winning the award have Stella Tiana Stegmann (47.6\,\%) and Marie Rauscher (46.6\,\%). Three other candidates (Lena Klahr, Kamila Joanna, and Marie Czuczman) have single-digit estimated probabilities to win, while the odds of winning the award are insignificant for the remaining seven candidates. These candidates have either been featured previously in other editions of the magazine, are non-natives, or do not have an Instagram account. In all historical cases investigated, candidates without an Instagram account, previous Playmates, or non-natives never won the award. From the candidates with positive odds, Marie Rauscher has the largest number of followers and a high level of interactions with her social media posts.


\vspace{-3ex}

\section{Conclusion}
\vspace{-2ex}

In this study, we investigated the predictability of the \textit{Playboy} magazine's \textit{Playmate of the Year} award using social media data from the candidates Instagram profiles. The study contributes to the fast-rising scholarly literature on quantification of success in different fields of human endeavour.
The findings described here provide some evidence for a relation between the social media activity and popularity on the one hand, and the contest's outcome on the other. 

However, the results presented here are based on a limited amount of data. Therefore, we were limited to a very parsimonious model including only a few variables. A larger data set would allow to measure other potentially influential effects, such as networks of followers to the magazine's Instagram account itself. These users might be more likely to participate in the polls and have an influence the outcome. In order to generalise the findings about the potential effect of social media marketing in contests, it would be beneficial to investigate other types of popular culture competitions.

Regardless, in providing first evidence of social media related success drivers in the Playboy's \textit{Playmate of the Year} contest, this short paper quantitatively assesses the relation between 'self-branding' on social media and success in a popularity competition.

\begin{table}[h!] \centering 
  \caption{Predictions of the German \textit{2020 Playboy Playmate of the Year} election.} 
  \label{tab:1} 
\begin{tabular}{cccccc} 
\toprule
 Playmate & Instagram\hspace{2ex} & Instagram\hspace{2ex} & Comments &  Estimated & Estim. \\ 
 & followers (tsd.)\hspace{2ex} & posts\hspace{2ex} & per post & probability (\%) & rank\\
\midrule
\rowcolor{gray!6}\rule{1ex}{0ex}Stella T. Stegmann & 16.9 & 234 & 8.8 & $47.6$ & $1$ \\ 
\rule{1ex}{0ex}Marie Rauscher & 22.5 & 265 & 15.3&  $46.6$ & $2$ \\ 
\rowcolor{gray!6}Lena Klahr & 17.3 & 264& ---&$4.4$ & $3$\\
\rule{1ex}{0ex}Kamila Joanna & 4.4 & 182 & 8.9&  $1.3$ & $4$ \\ 
\rowcolor{gray!6}\rule{1ex}{0ex}Marie Czuczman & 17.7 & 537 & 6.5 &  $0.1$ & $5$ \\ 
\rule{1ex}{0ex}Michelle Weisstuch & --- & --- & --- & $<0.01$ & $6-12$ \\ 
\rowcolor{gray!6}\rule{1ex}{0ex}Chelsie Aryn & 157 & 1,317& --- & $<0.01$ & $6-12$ \\ 
\rule{1ex}{0ex}Lena Bednarskaa &18.6 & 185 & 7.7 & $<0.01$ & $6-12$ \\ 
\rowcolor{gray!6}\rule{1ex}{0ex}Yvette Holleman &9.1& 100& 6.4 & $<0.01$ & $6-12$ \\ 
\rule{1ex}{0ex}Julia Logacheva &145& 627 & 14.8 & $<0.01$ & $6-12$ \\ 
\rowcolor{gray!6}\rule{1ex}{0ex}Katia Martin &87.3& 57& 13.5 & $<0.01$ & $6-12$ \\ 
\rule{1ex}{0ex}Tiffany van Roest &4.1& 122& --- & $<0.01$ & $6-12$ \\ 
\bottomrule
\end{tabular} 
\end{table}

\end{document}